\documentclass[
showkeys,12pt,
preprint,preprintnumbers,nofootinbib,
groupedaddress,superscriptaddress,amsmath,amssymb]{revtex4}
\usepackage{graphicx}
\usepackage{dcolumn}
\usepackage{bm}
\usepackage{amssymb}
\usepackage{amsmath}
\usepackage{epsfig}    
\usepackage{color}
\usepackage{slashed}
\usepackage{hhline}

\def\be{\begin{equation}}
\def\ee{\end{equation}}
\newcommand{\bea}{\begin{eqnarray}}
\newcommand{\eea}{\end{eqnarray}}
\newcommand{\nn}{\nonumber}

\numberwithin{equation}{section}

\begin{document}

\title{Radiative Neutrino Model with Inert Triplet Scalar}
%
\author{Hiroshi Okada}
\email{macokada3hiroshi@cts.nthu.edu.tw}
\affiliation{Physics Division, National Center for Theoretical Sciences, Hsinchu, Taiwan 300}

\author{Yuta Orikasa}
\email{orikasa@kias.re.kr}
\affiliation{School of Physics, KIAS, Seoul 130-722, Korea}
\affiliation{Department of Physics and Astronomy, Seoul National University, Seoul 151-742, Korea}

\date{\today}

\begin{abstract}
We study a one-loop induced radiative neutrino model with an inert isospin triplet scalar field in the general framework of $U(1)_Y$, in which we discuss current neutrino oscillation data, lepton flavor violations, muon anomalous magnetic moment, and a dark matter candidate depending on the number of hypercharges. We show global analysis combining all the constraints, and discuss the model.
\end{abstract}
\maketitle
\newpage

\section{Introduction}
Radiative seesaw models are well known as one of the economical and simultaneous explanations of tiny neutrino masses and dark matter (DM) due to having the strong correlations between them, and these issues are surely discussed beyond the standard model (SM).
Even though a vast literature has recently arisen in Refs.~\cite{Zee, Cheng-Li, zee-babu, Krauss:2002px, Ma:2006km, Aoki:2008av, Gustafsson:2012vj, Hambye:2006zn, Gu:2007ug, Sahu:2008aw, Gu:2008zf, Babu:2002uu, AristizabalSierra:2006ri, AristizabalSierra:2006gb,
Nebot:2007bc, Bouchand:2012dx, Kajiyama:2013sza,McDonald:2013hsa, Ma:2014cfa, Schmidt:2014zoa, Herrero-Garcia:2014hfa,
Ahriche:2014xra,Long1, Long2, Aoki:2010ib, Kanemura:2011vm, Lindner:2011it, 
Kanemura:2011jj, Aoki:2011he, Kanemura:2011mw,
Schmidt:2012yg, Kanemura:2012rj, Farzan:2012sa, Kumericki:2012bf, Kumericki:2012bh, Ma:2012if, Gil:2012ya, Okada:2012np,
Hehn:2012kz, Baek:2012ub, Dev:2012sg, Kajiyama:2012xg, Kohda:2012sr, Aoki:2013gzs, Kajiyama:2013zla, Kajiyama:2013rla, Kanemura:2013qva,Law:2013saa, Dasgupta:2013cwa, Baek:2013fsa, Okada:2014vla, Ahriche:2014cda, Ahriche:2014oda,Chen:2014ska,
Kanemura:2014rpa, Okada:2014oda, Fraser:2014yha, Okada:2014qsa, Hatanaka:2014tba, Baek:2015mna, Jin:2015cla,
Culjak:2015qja, Okada:2015nga, Geng:2015sza, Okada:2015bxa, Geng:2015coa, Ahriche:2015wha, Restrepo:2015ura, Kashiwase:2015pra, Nishiwaki:2015iqa, Wang:2015saa, Okada:2015hia, Ahriche:2015loa,
Ahn:2012cg, Ma:2012ez, Kajiyama:2013lja, Hernandez:2013dta, Ma:2014eka, Aoki:2014cja, Ma:2014yka, Ma:2015pma, 
Ma:2013mga, radlepton1, radlepton2, Okada:2014nsa, Brdar:2013iea, Okada:2015nca, Okada:2015kkj, Fraser:2015mhb, Fraser:2015zed, Adhikari:2015woo, Kanemura:2015cca, 
Bonnet:2012kz,Sierra:2014rxa, 
Davoudiasl:2014pya, Lindner:2014oea,Okada:2014nea, MarchRussell:2009aq, King:2014uha, Mambrini:2015sia, Baek:2014qwa}, there are few papers including inert isospin triplet scalar fields~\cite{Brdar:2013iea, Okada:2015nca}.
Especially, $1$-hypercharge triplet scalars (even) with nonzero vacuum expectation value (VEV) provides the type-II seesaw mechanism that naturally explains the tiny neutrino masses due to the smallness of VEV, which is  experimentally required.~\footnote{The theoretical reason has also been proposed by, {\it i.e.}, the paper in Ref.~\cite{Kanemura:2012rj}. }   
It also gives a lot of phenomenological aspects and/or constraints such as lepton flavor violation (LFV) processes through charged and/or neutral bosons, electroweak precision test, and the lowest bounds on their masses at large hadron collider (LHC).   

In this paper, 
we study a one-loop induced radiative neutrino model with an inert isospin triplet boson (instead of nonzero VEV), in which we  discuss
neutrino oscillations, lepton flavor violations, anomalous magnetic moment, and a (singlet-like) bosonic dark matter candidate to explain the relic density and the direct detection. Then we show  the several results from the global numerical analysis.
Here we introduce the simplest discrete symmetry $Z_2$ as an additional symmetry, which differentiates between SM fields and new fields and assures the stability of DM.


This paper is organized as follows.
In Sec.~II, we show our model,  including neutrino sector, LFVs, muon anomalous magnetic moment.
In Sec.~III, we analyze bosonic DM candidate to explain relic density and direct detection.
In Sec.~IV, we have a numerical analysis, and show some results.
We conclude and discuss in Sec.~V.


\section{ Model setup}
 \begin{widetext}
\begin{center} 
\begin{table}
\begin{tabular}{|c||c|c|c||c|c|c|}\hline\hline  
&\multicolumn{3}{c||}{Lepton Fields} & \multicolumn{3}{c|}{Scalar Fields} \\\hline
& ~$L_L$~ & ~$e_R^{}$~ & ~$L'$ ~ & ~$\Phi$~  & ~$\Delta$~ & ~$S^{-1+m}$ \\\hline 
$SU(2)_L$ & $\bm{2}$  & $\bm{1}$  & $\bm{2}$ & $\bm{2}$ & $\bm{3}$ & $\bm{1}$ \\\hline 
$U(1)_Y$ & $-\frac12$ & $-1$  & $-\frac{N}{2}$ & $\frac12$ & $\frac{1+N}{2}$ & $\frac{-1+N}{2}$  \\\hline
\end{tabular}
\caption{Contents of fermion and scalar fields
and their charge assignments under $SU(2)_L\times U(1)_Y$, where $m\equiv \frac{1+N}{2}$ is the quantum number of the electric charge.}
\label{tab:1}
\end{table}
\end{center}
\end{widetext}

In this section, we explain our model. 
The particle contents and their charges are shown in Tab.~\ref{tab:1}.
We add {three (or two)} iso-spin doublet vector-like exotic fermions $L'$ with $-N/2$ hypercharge,
an isospin triplet scalar $\Delta$  with $(1+N)/2$ hypercharge, and an isospin singlet scalar $S$  with $(-1+N)/2$ hypercharge to the SM, where $N$ is the odd inters. Here 
$S$ can be a real field for brevity, while $\Delta$ has to be a complex field.
We assume that  only the SM Higgs $\Phi$ have vacuum
expectation value (VEV), which is symbolized by $v/\sqrt2$. 

The relevant Lagrangian and Higgs potential under these symmetries are given by
\begin{align}
-\mathcal{L}_{Y}
&=
(y_{\ell})_{ij} \bar L_{Li} \Phi e_{Rj}  + (y_L)_{ij} \bar L_{L_i} L'_{R_j} S
+(y_{\Delta})_{ij} \bar L'^c_{L_i}(i\tau_2) \Delta L_{L_j}  + { (M_{L})_{ij}} \bar L'_{Li} L'_{Rj} + {\rm h.c.}, \nn\\
{\cal V}&=
m^2_{\Phi} \Phi^\dag\Phi + {m^2_{S_2}} S^2 + m^2_\Delta {\rm Tr}[\Delta^\dag\Delta]
\nn\\
&+  \lambda_0(\Phi^T (i\tau_2) \Delta^\dag \Phi S+{\rm.h.c.}) 
+ \lambda_{\Phi} |\Phi^\dag\Phi|^2  + \lambda_{S} S^4 + \lambda_{\Delta} [{\rm Tr}(\Delta^\dag\Delta)]^2+ \lambda_{\Delta}' {\rm Det}[\Delta^\dag\Delta]\nn\\
&+\lambda_{\Phi S}(\Phi^\dag\Phi)S^2 + \lambda_{\Phi\Delta}(\Phi^\dag\Phi) {\rm Tr}[\Delta^\dag\Delta]
 + \lambda'_{\Phi\Delta}\sum_{i=1}^3(\Phi^\dag\tau_i\Phi){\rm Tr}[\Delta^\dag\tau_i\Delta],
\label{Eq:lag-flavor}
\end{align}
where $i=1-3$, $j=1-3$, $\tau_i(i=1-3)$ is Pauli matrix, and the first term of $\mathcal{L}_{Y}$ can generates the SM
charged-lepton masses $m_\ell\equiv y_\ell v/\sqrt2$ after the electroweak spontaneous breaking of $\Phi$.
We work on the basis where all the coefficients are real and positive for simplicity. 
The scalar fields can be parameterized as 
\begin{align}
&\Phi =\left[
\begin{array}{c}
w^+\\
\frac{v+\phi+iz}{\sqrt2}
\end{array}\right],\quad 
\Delta =\left[
\begin{array}{cc}
\frac{\Delta^m}{\sqrt2} & \Delta^{1+m}\\
\Delta^{-1+m} & -\frac{\Delta^m}{\sqrt2}
\end{array}\right],
\label{component}
\end{align}
where $m\equiv \frac{1+N}{2}$ is the quantum number of the electric charge, $v~\simeq 246$ GeV is VEV of the Higgs doublet, and $w^\pm$
and $z$ are respectively GB 
which are absorbed by the longitudinal component of $W$ and $Z$ boson.
Inserting the tadpole condition; $\partial\mathcal{V}/\partial\phi|_{v}=0$,
the SM Higgs mass is given by $\sqrt{2\lambda_\Phi}v$.
While the resulting mass eigenstate and the matrix for the inert CP even boson masses $M_H(\Delta_R,S)$ are respectively given as \begin{align}
&O^TM_H(\Delta^{-1+m},S^{-1+m}) O = \left[\begin{array}{cc} m^2_{H_1^{-1+m}} & 0 \\ 0 & m^2_{H_2^{-1+m}} \end{array}\right],\\
&\left[\begin{array}{c} \Delta^{-1+m} \\ S^{-1+m} \end{array}\right] = 
O \left[\begin{array}{c} H_1^{-1+m} \\ H_2^{-1+m} \end{array}\right]=
\left[\begin{array}{cc} \cos\alpha & \sin\alpha \\ -\sin\alpha & \cos\alpha \end{array}\right]
\left[\begin{array}{c} H_1^{-1+m} \\ H_2^{-1+m} \end{array}\right],
\end{align}
with
\begin{align}
M_H & =
\left[\begin{array}{cc} m_\Delta^2+\frac{(\lambda_{\Phi\Delta}+\lambda_{\Phi\Delta}')v^2}{2} & -\sqrt2\lambda_0 v^2 \\ -\sqrt2\lambda_0 v^2 & 2m^2_S+\lambda_{\Phi S}v^2 \end{array}\right],
\quad \sin2\alpha=\frac{2\sqrt2 \lambda_0 v^2}{m^2_{H_1^{-1+m}}-m^2_{H_2^{-1+m}}}.
\end{align}
The other mass eigenstates is respectively given by
\begin{align}
& m^2_{\Delta^{-1+m}}=m_\Delta^2+\frac{\lambda_{\Phi\Delta}v^2}{2},\\
&m^2_{\Delta^{-1+m}}=m_\Delta^2+\frac{(\lambda_{\Phi\Delta}-\lambda_{\Phi\Delta}')v^2}{2}.
\end{align}

\subsection{ Neutrino mass matrix}
\begin{figure}[tb]
\begin{center}
\includegraphics[scale=0.45]{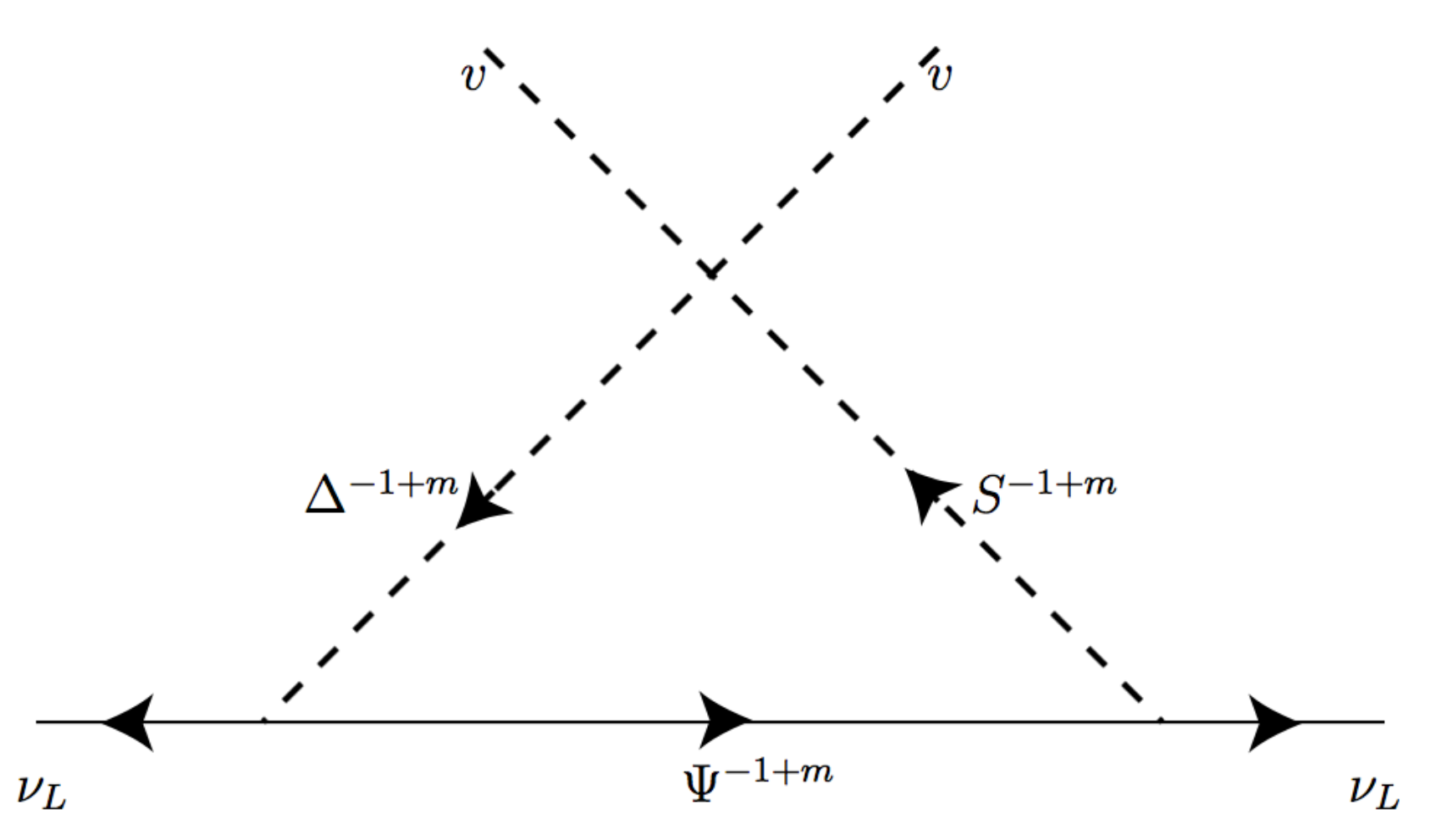}
\includegraphics[scale=0.45]{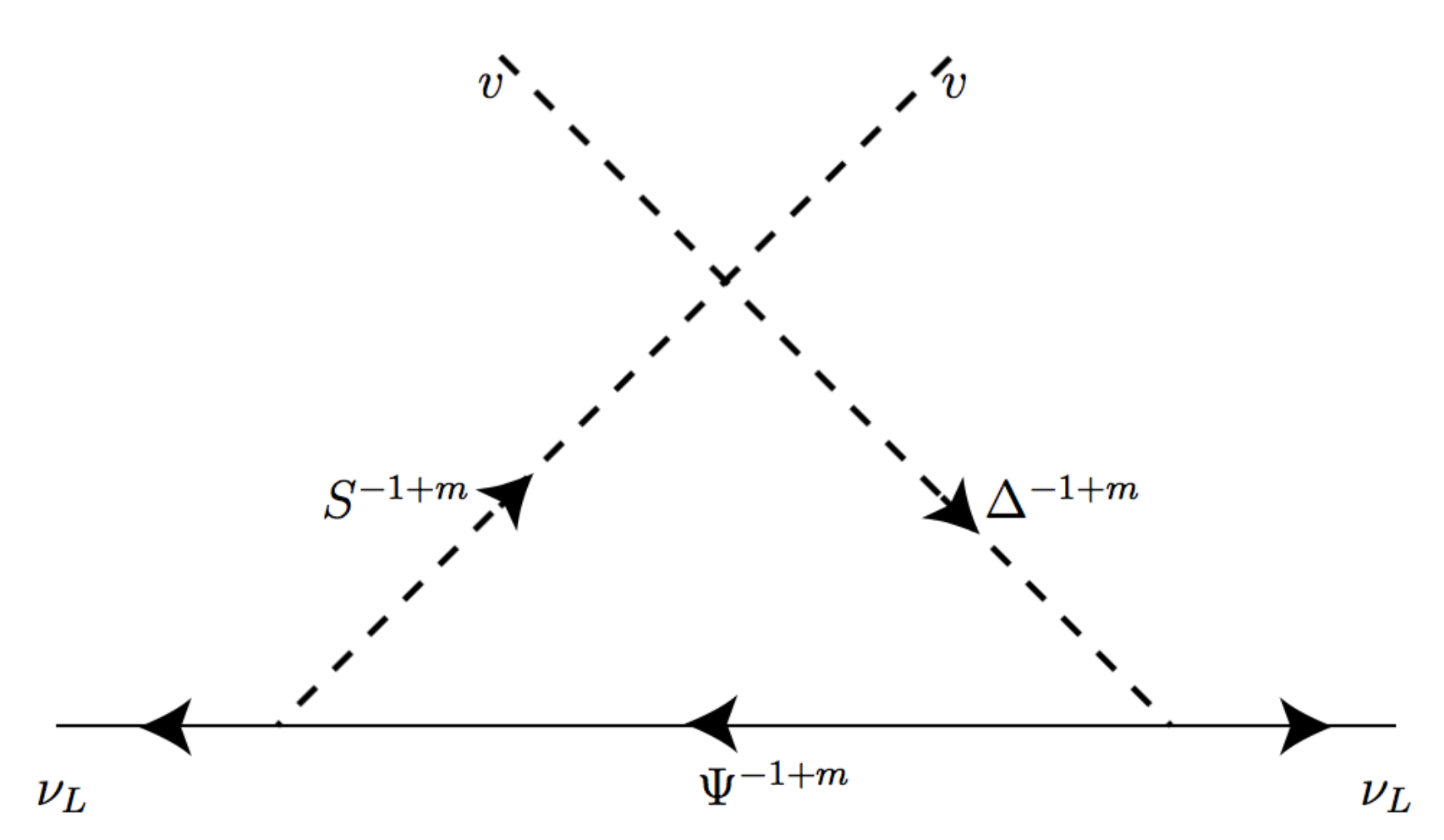}
\caption{ Neutrino masses at the one-loop level.
}   \label{fig:neut1}
\end{center}\end{figure}
At first we redefine relevant terms in terms of the mass eigenstate as
\begin{align}
\mathcal{L}_{Y}
&\supset
\frac{(y_L)_{ij} }{\sqrt2} \overline{ \Psi_{R_i}^{-1+m}}  \nu_{L_j} (-s_\alpha H_1^{-1+m} + c_\alpha H_2^{-1+m})\nn\\
&+
\frac{(y_\Delta)_{ij} }{\sqrt2} 
 \overline{ (\Psi^c_{L_i})^{1-m}}  \nu_{L_j} (-c_\alpha H_1^{-1+m} + s_\alpha H_2^{-1+m}),
\label{Eq:lag-mass}
\end{align}
where we define the isospin doublet exotic fermion as $L'\equiv [\Psi^{1-m},\Psi^{-m}]^T$.

Then the dominant contribution to the active neutrino mass matrix $m_\nu$  is given at one-loop level as shown in Figure~\ref{fig:neut1}, and its formula is given by 
\begin{align}
&(m_{\nu})_{ab}
=-\frac{s_\alpha c_\alpha}{2(4\pi)^2}
\sum_{}[(y_L)_{ai}(y_\Delta)_{ib} + (y_L)_{bi}(y_\Delta)_{ia}] M_{L_i}
\left[
\frac{X_{1,i} }{X_{1,i}-1}\ln[X_{1,i}] - \frac{X_{2,i} }{X_{2,i}-1}\ln[X_{2,i}]
 \right],
\end{align}
where we define $X_{a,i}\equiv (m_{H_a^{-1+m}}/M_{L_i})^2$ (a=1,2).
One finds that the structure of this formula is the same as Ma model~\cite{Ma:2006km}.
 $({m}_\nu)_{ab}$ can be generally diagonalized by the Pontecorvo-Maki-Nakagawa-Sakata mixing matrix $V_{\rm MNS}$ (PMNS)~\cite{Maki:1962mu} as
\begin{align}
({m}_\nu)_{ab} &=(V_{\rm MNS} D_\nu V_{\rm MNS}^T)_{ab},\quad D_\nu\equiv (m_{\nu_1},m_{\nu_2},m_{\nu_3}),
\\
V_{\rm MNS}&=
\left[\begin{array}{ccc} {c_{13}}c_{12} &c_{13}s_{12} & s_{13} e^{-i\delta}\\
 -c_{23}s_{12}-s_{23}s_{13}c_{12}e^{i\delta} & c_{23}c_{12}-s_{23}s_{13}s_{12}e^{i\delta} & s_{23}c_{13}\\
  s_{23}s_{12}-c_{23}s_{13}c_{12}e^{i\delta} & -s_{23}c_{12}-c_{23}s_{13}s_{12}e^{i\delta} & c_{23}c_{13}\\
  \end{array}
\right],
\end{align}
where we neglect the Majorana phase as well as Dirac phase $\delta$ in the numerical analysis for simplicity.
The following neutrino oscillation data at 95\% confidence level~\cite{pdf} is given as
\begin{eqnarray}
&& 0.2911 \leq s_{12}^2 \leq 0.3161, \; 
 0.5262 \leq s_{23}^2 \leq 0.5485, \;
 0.0223 \leq s_{13}^2 \leq 0.0246,  
  \\
&& 
  \ |m_{\nu_3}^2- m_{\nu_2}^2| =(2.44\pm0.06) \times10^{-3} \ {\rm eV}^2,  \; 
  \ m_{\nu_2}^2- m_{\nu_1}^2 =(7.53\pm0.18) \times10^{-5} \ {\rm eV}^2, \nn
  \label{eq:neut-exp}
  \end{eqnarray}
where we assume one of three neutrino masses is zero with normal ordering in our analysis below.
%
The observed PMNS matrix can be realized by introducing the following parametrization.
Here we can parametrize the Yukawa coupling $y_L$ as follows;
\begin{align}
y_L 
&=\frac12 (V_{\rm MNS}^* D_\nu V_{\rm MNS}^\dag+ A) y_\Delta^{-1}R^{-1},
\label{yl-sol}
\\
R_{ii}&\equiv \frac{s_\alpha c_\alpha M_{L_i}}{2(4\pi)^2}
\left[
\frac{X_{1,i} }{X_{1,i}-1}\ln[X_{1,i}] - \frac{X_{2,i} }{X_{2,i}-1}\ln[X_{2,i}]
 \right],\label{R-sol}
\end{align}
where $A$ is an arbitrary anti-symmetric matrix with complex values.
In the numerical analysis as can be discussed later, we determine the value of $y_L$ to make a random plot for $y_\Delta$, $R$, and $A$.

\subsection{ Lepton Flavor Violations}
\label{lfv-lu}
$\ell_b\to\ell_a \gamma$ processes arise from the following term via one-loop diagrams
\begin{align}
\mathcal{L}_{Y}
&\supset
-\frac{(y_L)_{ij} }{\sqrt2} \overline {\Psi_{R_i}^m}  \ell_{L_j} (-s_\alpha H_1^{-1+m} + c_\alpha H_2^{-1+m})
+
{(y_\Delta)_{ij} } 
\overline {\Psi_{L_i}^m}
 \ell_{L_j} \Delta^{1+m}
+
\frac{(y_\Delta)_{ij} }{\sqrt{2}} 
\overline {\Psi_{R_i}^{-1+m}}
\ell_{L_j} \Delta^{m}.
\label{Eq:lag-mass}
\end{align}
Then the branching ratio of ${\rm BR}(\ell_b\to\ell_a \gamma)$ is defined by
\begin{align}
{\rm BR}(\ell_b\to\ell_a \gamma)
=
\frac{48\pi^3 C_b\alpha_{\rm em}}{{\rm G_F^2} m_b^2 }(|a_R|^2+|a_L|^2),
\end{align}
where $\alpha_{\rm em}\approx1/137$ is the fine-structure constant,
$C_{b}=(1,1/5)$ for ($b=\mu,\tau$), ${\rm G_F}\approx1.17\times 10^{-5}$ GeV$^{-2}$ is the Fermi constant, $a_L$ and $a_R$ are respectively computed as
\begin{align}
(a_R)_{ab}&=-\frac{m_{\ell_b}}{(4\pi)^2}\int dxdy dz\delta(x+y+z-1)xy \nn\\
&
\sum_{i=1}^3\left[
\frac{m(y^\dag_L)_{ai} (y_L)_{ib} }{2}
\left( \frac{s^2_\alpha}{\Delta[m_{H_1^{-1+m}},M_{L_i}]}+ \frac{c^2_\alpha}{\Delta[m_{H_2^{-1+m}},M_{L_i}]} \right)\right.\nn\\
&\left.
+ \frac{(m-1)(y^\dag_L)_{ai} (y_L)_{ib} }{2}
\left( \frac{s^2_\alpha}{\Delta[M_{L_i},m_{H_1^{-1+m}}]}+ \frac{c^2_\alpha}{\Delta[M_{L_i},m_{H_2^{-1+m}}]} \right)\right.\nn\\
&\left.
-(y^\dag_\Delta)_{ai}(y_\Delta)_{ib}
\left( \frac{1+m}{\Delta[M_{L_i}, m_{\Delta^{1+m}}]}+ \frac{m}{\Delta[m_{\Delta^{1+m}},M_{L_i}]} 
+ \frac{m}{\Delta[M_{L_i},m_{\Delta^{m}}]} + \frac{m-1}{\Delta[m_{\Delta^{m}},M_{L_i}]}  \right)
\right],\\
(a_L)_{ab}&=-\frac{m_{\ell_a}}{(4\pi)^2}\int dxdy dz\delta(x+y+z-1)xz \nn\\
&
\sum_{i=1}^3\left[
\frac{m(y^\dag_L)_{ai} (y_L)_{ib} }{2}
\left( \frac{s^2_\alpha}{\Delta[m_{H_1^{-1+m}},M_{L_i}]}+ \frac{c^2_\alpha}{\Delta[m_{H_2^{-1+m}},M_{L_i}]} \right)\right.\nn\\
&\left.
+ \frac{(m-1)(y^\dag_L)_{ai} (y_L)_{ib} }{2}
\left( \frac{s^2_\alpha}{\Delta[M_{L_i},m_{H_1^{-1+m}}]}+ \frac{c^2_\alpha}{\Delta[M_{L_i},m_{H_2^{-1+m}}]} \right)\right.\nn\\
&\left.
-(y^\dag_\Delta)_{ai}(y_\Delta)_{ib}
\left( \frac{1+m}{\Delta[M_{L_i}, m_{\Delta^{1+m}}]}+ \frac{m}{\Delta[m_{\Delta^{1+m}},M_{L_i}]} 
+ \frac{m}{\Delta[M_{L_i},m_{\Delta^{m}}]} + \frac{m-1}{\Delta[m_{\Delta^{m}},M_{L_i}]}  \right)
\right],\\
& \Delta[m_\rho,m_\sigma]\approx 
x m^2_\rho +(y+z)m^2_\sigma,
\end{align} 
where $M_{\Psi^{-1+m}}=M_{\Psi^m}=M_L$, and if $m_{\ell_a}\approx 0$, one can approximate the above formula to be
\begin{align}
{\rm BR}(\ell_b\to\ell_a \gamma)
\approx \frac{48\pi^3 C_b\alpha_{\rm em}}{{\rm G_F}^2 m_{\ell_b}^2}|(a_R)_{ab}|^2.
\end{align}
Notice here that three body decay processes $\ell_a\to\ell_b\ell_c\ell_d$ at the one-loop box type of diagrams are negligible comparing to the ${\rm BR}(\ell_b\to\ell_a \gamma)$ types of LFVs~\cite{Toma:2013zsa}.

\begin{table}[t]
\begin{tabular}{c|c|c} \hline
Process & $(b,a)$ & Experimental bounds ($90\%$ CL) \\ \hline
$\mu^{-} \to e^{-} \gamma$ & $(2,1)$ &
	$\text{Br}(\mu \to e\gamma) < 5.7 \times 10^{-13}$  \\
$\tau^{-} \to e^{-} \gamma$ & $(3,1)$ &
	$\text{Br}(\tau \to e\gamma) < 3.3 \times 10^{-8}$ \\
$\tau^{-} \to \mu^{-} \gamma$ & $(3,2)$ &
	$\text{Br}(\tau \to \mu\gamma) < 4.4 \times 10^{-8}$  \\ \hline
\end{tabular}
\caption{Summary of $\ell_b \to \ell_a \gamma$ process and the lower bound of experimental data~\cite{Adam:2013mnn}.}
\label{tab:Cif}
\end{table}

\subsection{$\mu-e$ conversion}
The $\mu-e$ conversion rate $R$ 
 is given by~\cite{Hisano:1995cp}
\begin{align}
R&=\frac{\Gamma(\mu\to e)}{\Gamma_{\rm capt}},\\
\Gamma(\mu\to e)&=
C_{\mu e}
\left[
\left|Z\left(b_L^\gamma-\frac{a_R}{m_{\ell_b}}\right)
- b_L^Z \frac{(2Z+N)A_u+(Z+2N)A_d}{2(s_{tw}c_{tw})^2}  \right|^2
+\left|\frac{Z a_L}{m_{\ell_a}}\right|^2
\right]_{b=\mu,a=e},
\end{align}
where $C_{\mu e}\equiv4\alpha_{\rm em}^5 \frac{Z^4_{\rm eff}|F(q)|^2 m^5_\mu}{Z}$, $A_u\equiv -\frac12-\frac43s_{tw}^2$, $A_d\equiv -\frac12+\frac23s_{tw}^2$, $\sin^2\theta_w\equiv s_{tw}^2\approx0.23$. The values of $\Gamma_{\rm capt}$, $Z$, $N$, $Z_{\rm eff}$, and $F(q)$ depend on the kind of nuclei.
But  we use Titanium, because its sensitivity will be improved by several orders of magnitude~\cite{Hungerford:2009zz, Cui:2009zz} in near future compared to the current bound~\cite{Dohmen:1993mp};
\begin{align}
R_{\rm Ti}^{\rm Current \ bound} \lesssim 4.3\times 10^{-12}\to R_{\rm Ti}^{\rm Future \ bound} \lesssim 10^{-18}.
\end{align}
 Then these values are determined by $\Gamma_{\rm capt}=2.59\times10^6$ sec$^{-1}$, $Z=22$, $N=26$, $Z_{\rm eff}=17.6$,  and $|F(-m_\mu^2)|=0.54$~\cite{Alonso:2012ji}.

$b^V_L$ has to be determined by our model, and its formula is given by
\begin{align}
(b_L)^V_{ab}&=-\frac{1}{(4\pi)^2}\int dxdy dz\delta(x+y+z-1)z(1-z) \nn\\
&
\sum_{i=1}^3\left[
\frac{m(y^\dag_L)_{ai} (y_L)_{ib} }{2}
\left( \frac{s^2_\alpha}{\Delta[m_{H_1^{-1+m}},M_{L_i}]}+ \frac{c^2_\alpha}{\Delta^V[m_{H_2^{-1+m}},M_{L_i}]} \right)\right.\nn\\
&\left.
-(y^\dag_\Delta)_{ai}(y_\Delta)_{ib}
\left( \frac{m}{\Delta^V[m_{\Delta^{1+m}},M_{L_i}]} 
 + \frac{m-1}{\Delta^V[m_{\Delta^{m}},M_{L_i}]}  \right)
\right],\\
& \Delta^V[m_\rho,m_\sigma]\approx 
x m^2_\rho +(y+z)m^2_\sigma +z(x+z-1)m^2_V,
\end{align}
where $V\equiv (\gamma, Z)$, and $m_\gamma=0$, and $m_Z\approx 91.19$ GeV.~\footnote{We would like thank referee to point out this process that gives rather strong constraint to our model. }

\subsection{Muon anomalous magnetic moment}
The muon anomalous magnetic moment (muon $g-2$) has been 
measured at Brookhaven National Laboratory. 
The current average of the experimental results is given by~\cite{bennett}
\begin{align}
a^{\rm exp}_{\mu}=11 659 208.0(6.3)\times 10^{-10}. \notag
\end{align}
It has been well known that there is a discrepancy between the
experimental data and the prediction in the SM. 
The difference $\Delta a_{\mu}\equiv a^{\rm exp}_{\mu}-a^{\rm SM}_{\mu}$
was calculated in Ref.~\cite{discrepancy1} as 
\begin{align}
\Delta a_{\mu}=(29.0 \pm 9.0)\times 10^{-10}, \label{dev1}
\end{align}
and it was also derived in Ref.~\cite{discrepancy2} as
\begin{align}
\Delta a_{\mu}=(33.5 \pm 8.2)\times 10^{-10}. \label{dev2}
\end{align}
The above results given in Eqs. (\ref{dev1}) and (\ref{dev2}) correspond
to $3.2\sigma$ and $4.1\sigma$ deviations, respectively. 

Our formula of muon $g-2$ can simply be given by
\begin{align}
\Delta a_\mu\approx -\frac{m_\mu}{2}[{(a_R)_{22}+(a_L)_{22}}],\label{damu}
\end{align}
where the lower index $2$ of $a_{R(L)}$ represents the muon.

\section{Model features }
 \begin{widetext}
\begin{center} 
\begin{table}[tbc]
\begin{tabular}{|c||c|c|c||c|c|c|}\hline\hline  
&\multicolumn{3}{c||}{Lepton Fields} & \multicolumn{3}{c|}{Scalar Fields} \\\hline
& ~$L_L$~ & ~$e_R^{}$~ & ~$L'$ ~ & ~$\Phi$~  & ~$\Delta$~ & ~$S$ \\\hline 
$SU(2)_L$ & $\bm{2}$  & $\bm{1}$  & $\bm{2}$ & $\bm{2}$ & $\bm{3}$ & $\bm{1}$ \\\hline 
$U(1)_Y$ & $-1/2$ & $-1$  & $-1/2$ & $1/2$ & $1$ & $0$  \\\hline
$Z_2$ & $+$ & $+$   & $-$ & $+$& $-$  & $-$  \\\hline
\end{tabular}
\caption{Contents of fermion and scalar fields
and their charge assignments under $SU(2)_L\times U(1)_Y\times Z_2$.}
\label{tab:1}
\end{table}
\end{center}
\end{widetext}

\subsection{ The case of $m=1$}
The case of $m=1$ is  specific, because we accidentally have  some additional terms as follows:
\begin{align}
\bar L^c_L(i\tau_2)\Delta L_L, \quad \bar L'^c(i\tau_2)\Delta L', \quad 
\Phi^T (i\tau_2)\Delta^\dag \Phi,  \quad 
\bar L'_L L'_R S.
\end{align}  
Therefore the $m=1$ case cannot realize the general model by itself any longer as discussed in the previous section, since the neutrino masses are induced through the above first term at the tree level. 
Thus we impose an additional $Z_2$ symmetry for new fields as shown in Tab.~\ref{tab:1}.
As a result, all the terms are originated from the general Lagrangian, and
the $Z_2$ symmetry plays a role in assuring the stability of DM candidates $S$ or the neutral component of $\Delta$.
Here  $S$ is a real field for brevity, while $\Delta$ has to be a complex field. Then all the discussions are same as the general one except the DM candidates. So we will discuss about the DM candidates below.

{\it Relic density of Dark Matter}:
We have two DM candidates; the lightest of $N_i$ or $H_i$, where $N_i$ is the neutral component of $L'$. However since the fermionic candidate $N$ 
 interacts Z boson,  it is ruled out by the experiment of direct detection searches such as LUX~\cite{Akerib:2013tjd}.
 Hence we consider the bosonic DM candidate.
 Moreover, we identify DM as the isospin singlet-like boson $H_2$ to evade the constraint of S-T-U parameter~\cite{j.beringer}. 
 Notice here that $H_2$ is redefined by $X$, and its mass is symbolized by $M_X$ hereafter.
  
{\it Direct detection}: We have a spin independent scattering cross section with nucleon through the SM Higgs($\phi$) portal process and its form is simply given by
\begin{align}
\sigma_N\approx 0.082\frac{\mu_1^2 m_N^4}{\pi v^2M_X^2 m_\phi^4},
\end{align}
where $\mu_1\equiv -2(\lambda_{\Phi S}+2 \sqrt2 \sin\alpha \lambda_0)v$, and the mass of neutron, which is symbolized by $m_N$, is around 0.939 GeV. LUX suggests that $\sigma_N$ should be less than ${\cal O}$(10$^{-45}$) cm$^2$ at ${\cal O}$(10) GeV mass range of DM.

{\it Relic density}: Our relevant processes for the thermal averaged cross section comes from annihilations of $2X\to 2\phi$, $2X\to \nu_L\bar\nu_L$, and $2X\to \ell\bar\ell$,~\footnote{Since we assume to be $\alpha<<1$, any processes including gauge bosons $Z/W$ are suppressed by $\sin^2\alpha$ at least. Hence we can neglect these processes.} and their form is given by~\cite{Griest:1990kh, Edsjo:1997bg}
\begin{align}
&\sigma v_{\rm rel}\approx \int_0^{\pi} \sin\theta d\theta \frac{ |\bar M|^2}{16\pi s}\sqrt{1-\frac{4 m_f^2}{s}},
\end{align}
where
\begin{align}
 |\bar M|^2\approx  &|\bar M(2X\to 2\phi)|^2+ |\bar M(2X\to\nu_L\bar\nu_L)|^2+ |\bar M(2X\to\ell\bar\ell)|^2,\\ 
|\bar M(2X\to 2\phi)|^2&
\approx \frac12
\left|\frac{\mu_1}{v}-\frac{\mu_1\mu_2}{s-m^2_\phi} -\mu_1^2 \left(\frac{1}{t-m^2_{H_2}}+\frac{1}{u-m^2_{H_2}}\right) \right|^2,
\label{main-mode}
\\
|\bar M(2X\to\nu_L\bar\nu_L)|^2&
\approx 
2{\sum_{a,b}^{1-3}|Y_{i,b}|^2 |Y_{i,a}|^2}
\frac{M_X^6 (2M_X^2 + M_{N_b}^2+ M_{N_a}^2)}{(M_X^2 + M_{N_b}^2)^2(M_X^2 + M_{N_a}^2)^2}
v_{\rm rel}^2,\\
|\bar M(2X\to\ell\bar\ell)|^2&
\approx 
4 {\sum_{a,b}^{1-3}|(y_L)_{i,b}|^2 |(y_L)_{i,a}|^2\cos^4\alpha }
\frac{M_X^6 (2M_X^2 + M_{N_b}^2+ M_{N_a}^2)}{(M_X^2 + M_{N_b}^2)^2(M_X^2 + M_{N_a}^2)^2}
v_{\rm rel}^2,
\end{align}
where $Y_{i,j}\equiv(y_L)_{i,j}\cos\alpha+(y_\Delta)_{i,j}\sin\alpha$, and $\mu_2\equiv 6\lambda_\Phi v$.
Then the relic density is given by
\begin{align}
\Omega h^2\approx \frac{1.07\times10^9}
{g^{1/2}_* M_{\rm pl}[{\rm GeV}] \int_{x_f}^\infty \left(\frac{a_{\rm eff}}{x^2}+6\frac{b_{\rm eff}}{x^3} \right)},
\end{align}
where $g_*\approx 100$ is the total number of effective relativistic degrees of freedom at the time of freeze-out,
$M_{\rm pl}=1.22\times 10^{19}[{\rm GeV}] $ is Planck mass, $x_f\approx25$, and $ a_{\rm eff}$ and $ a_{\rm eff}$ are derived by
expanding $\sigma v_{\rm rel}$  in terms of $v_{\rm rel}$ up to $v_{\rm rel}^2$ as
\begin{align}
\sigma v_{\rm rel}\approx a_{\rm eff}+ b_{\rm eff} v^2_{\rm rel}.
\end{align}
The observed relic density reported by Planck suggest that $\Omega h^2\approx 0.12$~\cite{Ade:2013zuv}.

\begin{figure}[tb]
\begin{center}
\includegraphics[scale=0.6]{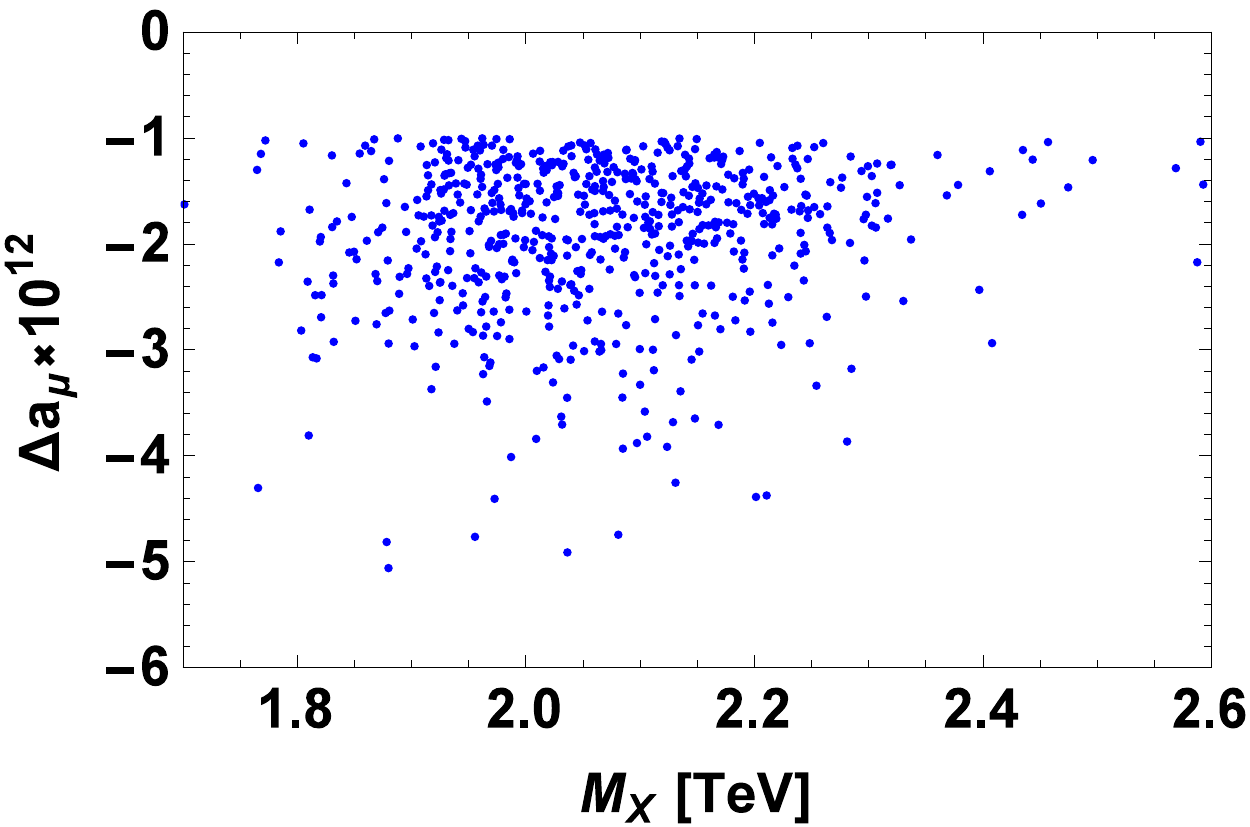}
\includegraphics[scale=0.6]{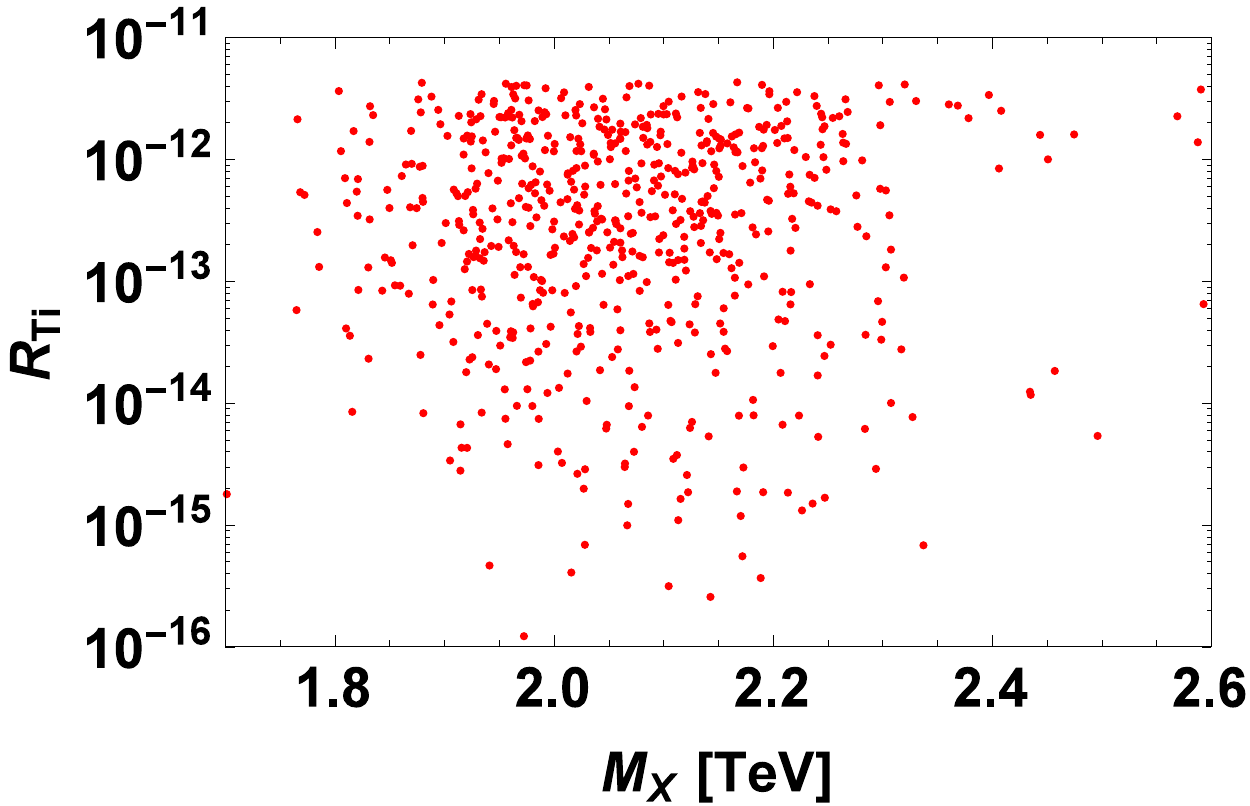}
\includegraphics[scale=0.6]{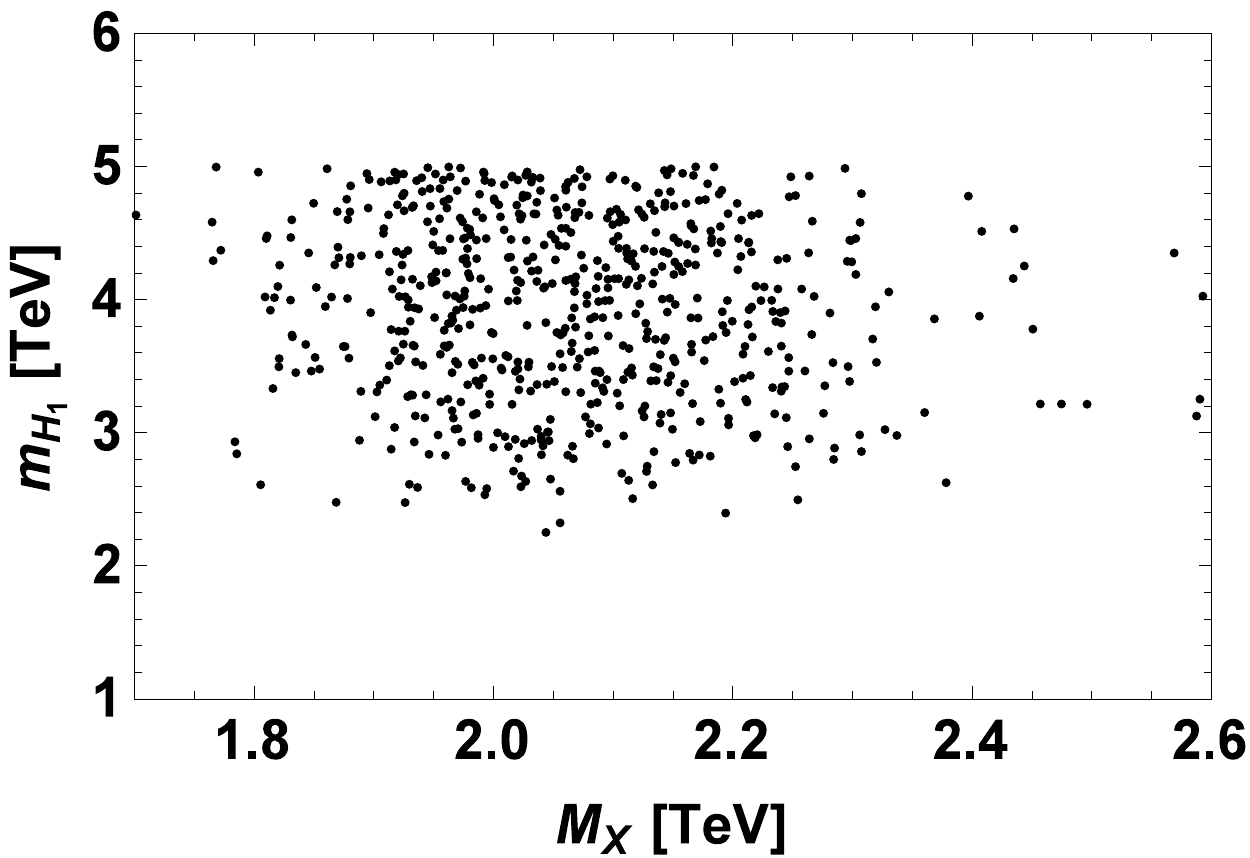}
\caption{Numerical results: The upper-left side figure represents the scattering plot in the DM mass and the muon anomalous magnetic moment. It suggests that the discrepancy is at most $|\Delta a_\mu|={\cal O}(10^{-12})$ with negative value that directly comes from the constraint of $\mu-e$ conversion rate. The allowed region of the DM mass comes from the measured relic density.
The upper-right side figure represents the scattering plot in the DM mass and the $\mu-e$ conversion rate with the target Ti that satisfies the current upper bound. Our minimal value of $R_{\rm Ti}$ is ${\cal O}(10^{-16})$, which can be tested in the future experiment that will reaches ${\cal O}(10^{-18})$.
The lower figure shows the scattering plot in the DM mass and the heavier inert scalar mass.
It suggests that its mass should be a few times as large as the mass of DM to evade LFVs of $\ell_b\to\ell_a \gamma$. 
 }
\label{results}
\end{center}
\end{figure}

\section{Numerical results}
In this section, we randomly select values of the twelve parameters within the corresponding ranges
\begin{align}
& M_X \in [0.1\,\text{TeV}, 3\,\text{TeV}],\quad
(M_E,\ M_N,\ m_{H_1},\ m_{\Delta^\pm},\ m_{\Delta^{\pm\pm}}) \in [M_X,\ 5\,\text{TeV}],\nn\\
& \delta \in [0, 2\pi],\quad \alpha \in [-0.3, 0.3],\quad
[(y_\Delta)_{i,j} ,A] \in [-1,1],\ {\rm for}\ (i,j)=(1,1),\ (1,3),\ (3,1),\ (3,3),  \nn\\ 
&[ (y_\Delta)_{2,1} ,\ (y_\Delta)_{2,3} , \ A] \in [-1,-0.5],\  [(y_\Delta)_{1,2} ,\ (y_\Delta)_{3,2} ]= A \in [0.5,1],\nn\\
&[ (y_\Delta)_{2,2} ,\ (y_\Delta)_{2,3}] = A \in [-0.1,0.1],\
\quad
[ \lambda_0 ,\  \lambda_{\Phi S}] \in [0, 1],
	\label{range_scanning}
\end{align}
to reproduce neutrino oscillation data,\footnote{In our analysis, we have used the normal ordering case. But we have checked that inverted ordering provides almost the same result.} LFVs, the constraint of the direct detection searches\footnote{We conservatively take the constraint $\sigma_N\lesssim 10^{-45}$cm$^2$ for all the mass region of DM.} and the observed relic density of DM.~\footnote{We take allowed region to be 0.11$\le\Omega h^2\le$0.13 instead of the exact value.} 
In this analysis, we are preparing 1 million sample points. 

\subsection{The case $m=1$}
Here we show numerical results  for $m=1$ case as shown in Figure~\ref{results}.
The upper-left figure represents the scattering plot in the DM mass and the muon anomalous magnetic moment.
It suggests that the discrepancy is at most $|\Delta a_\mu|={\cal O}(10^{-12})$ with negative value that directly comes from the constraint of $\mu-e$ conversion rate.
The upper-right side figure represents the scattering plot in the DM mass and the $\mu-e$ conversion rate with the target Ti that satisfies the current upper bound. Our minimal value of $R_{\rm Ti}$ is ${\cal O}(10^{-16})$, which can be tested in the future experiment that will reaches ${\cal O}(10^{-18})$.
The lower side figure shows the scattering plot in the DM mass and the heavier inert scalar mass.
It suggests that these masses should be a few TeV mass scale to evade LFVs and the allowed region of DM mass should be  around 
\begin{align} 1.7\ [{\rm TeV}] \le M_X\le 2.6\ [{\rm TeV}],\end{align}
 which is rather narrow comparing to our scanning  space.~\footnote{Notice here that there exists a solution at the resonant point of the half mass of the SM Higgs boson, although we do not take this point as our range. }
 This result directly stems from the constraint of the relic density, and the dominant mode is $2X\to2\phi$ in Eq.~(\ref{main-mode}) due to the s-wave. 
 


\begin{figure}[tbc]
\begin{center}
\includegraphics[scale=0.5]{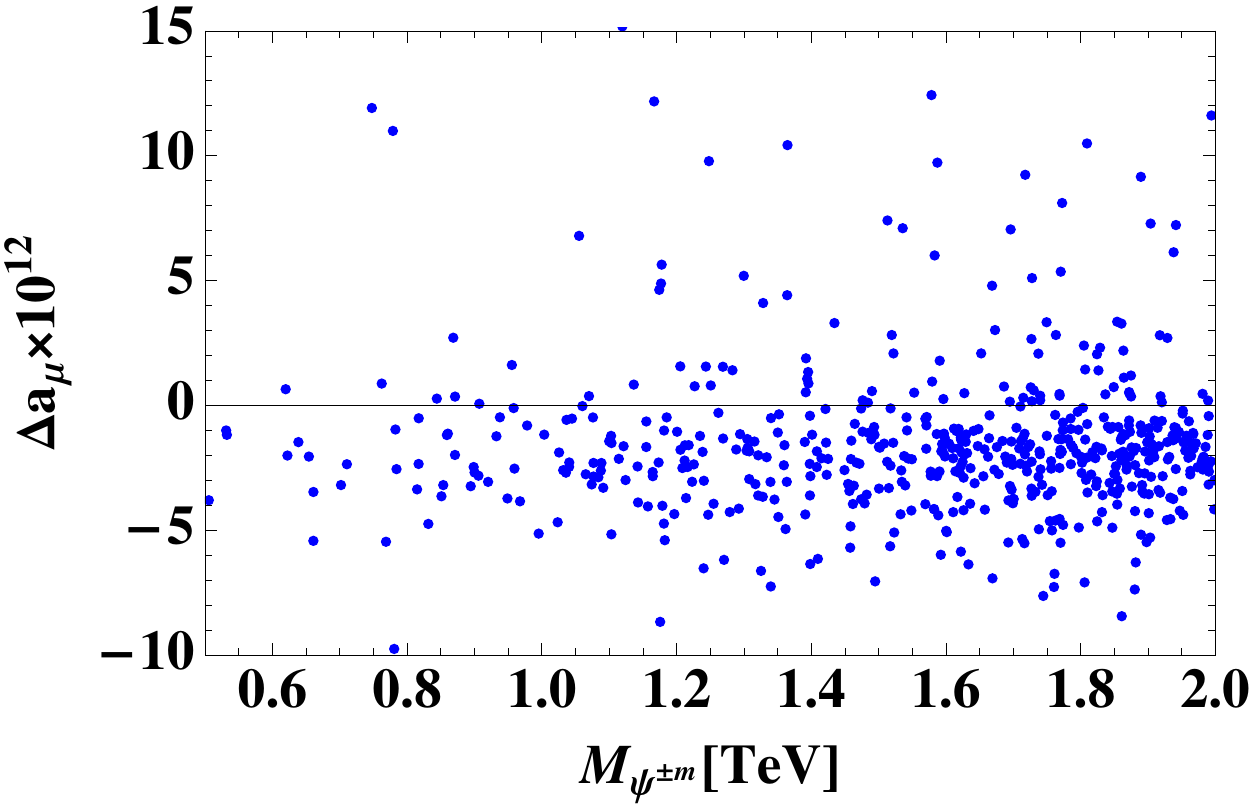}
\includegraphics[scale=0.5]{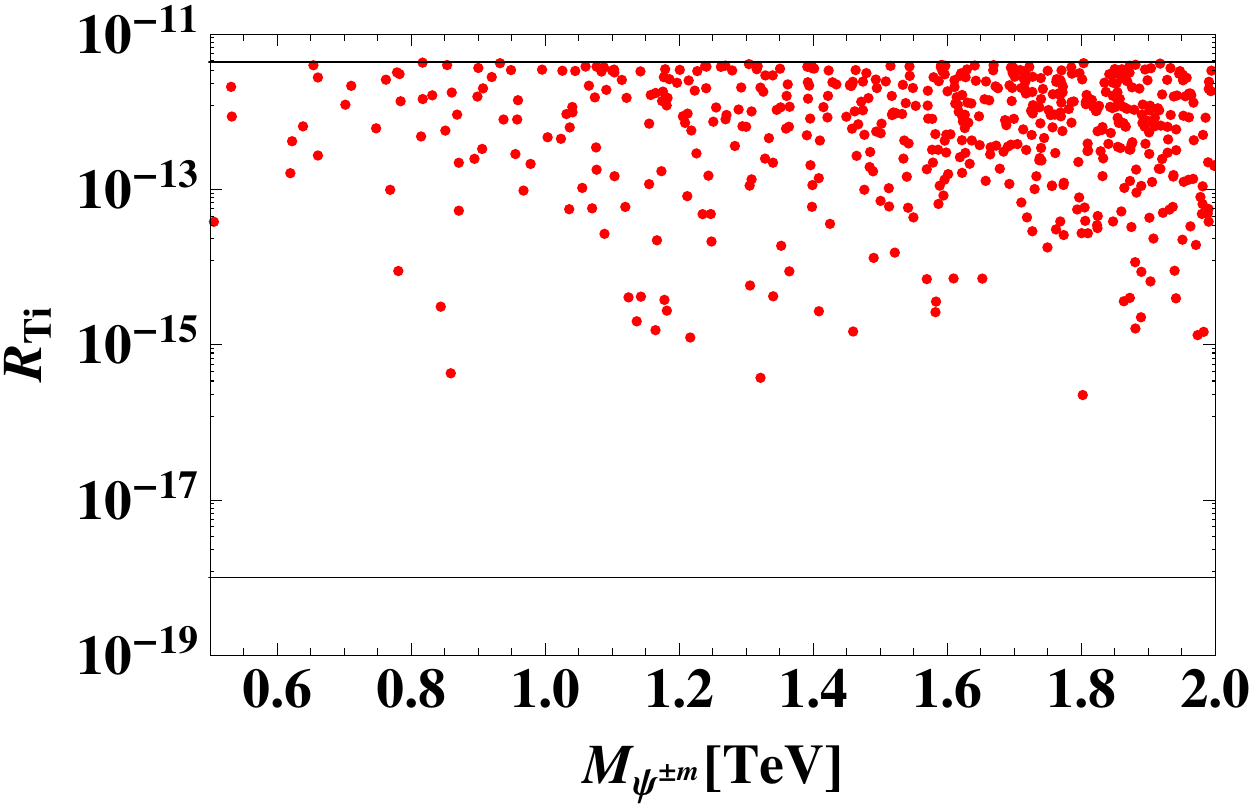}
\includegraphics[scale=0.5]{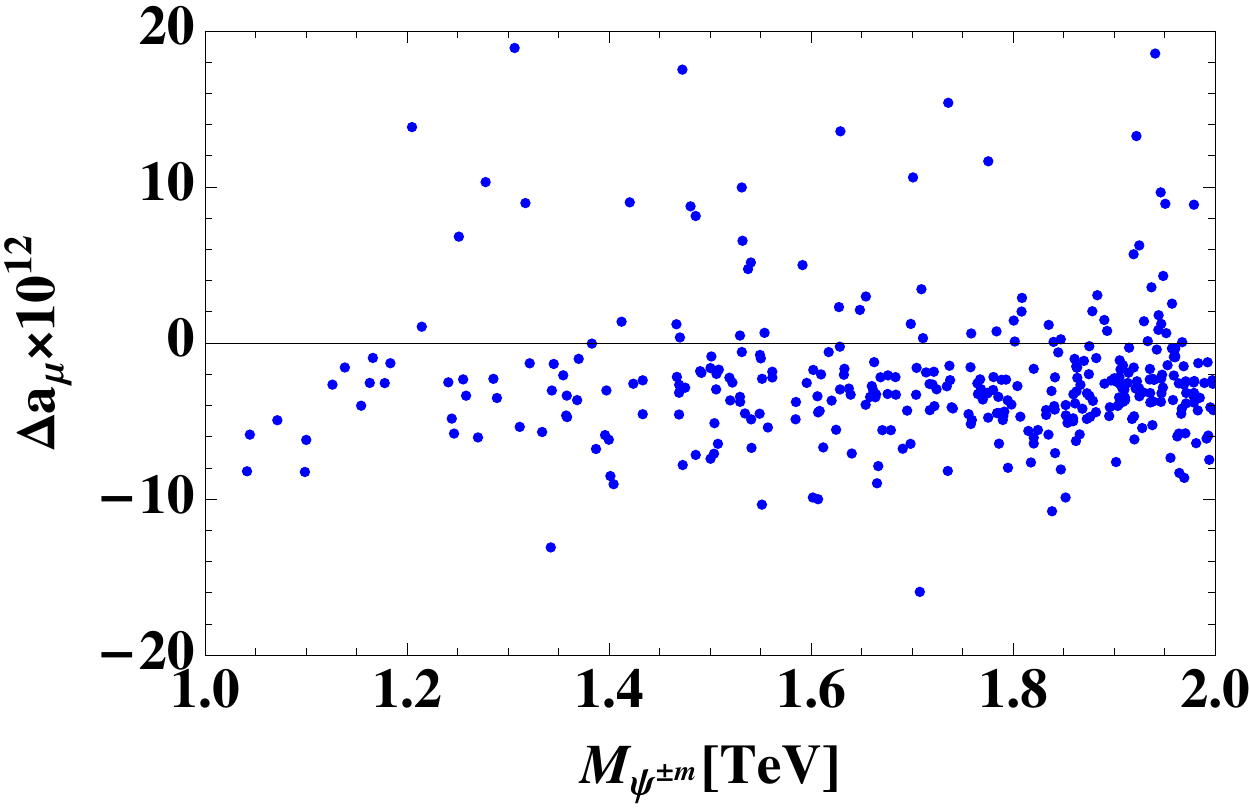}
\includegraphics[scale=0.5]{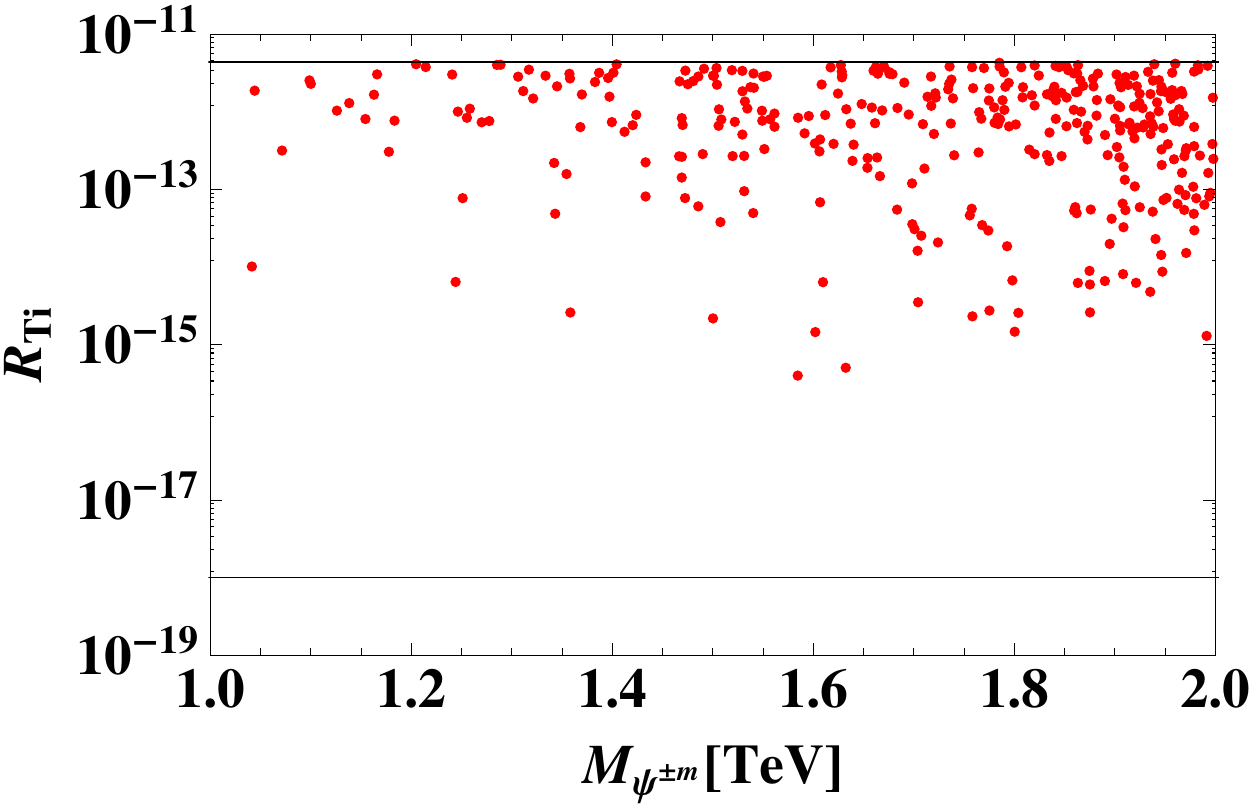}
\caption{The blue scattering plots represents the muon anomalous magnetic moment in terms of the exotic fermion ($\Psi^\pm$) mass , where the upper figure represents $m=2$, and the lower one does $m=3$. Here the neutrino oscillation data and the LFVs are taken into consideration as constraints.  
These suggest that the maximal values of discrepancy are respectively $\Delta a_\mu={\cal O}(1.5\times 10^{-11})$ and $\Delta a_\mu={\cal O}(2.0\times 10^{-11})$ that are below the current experimental value in  Eq.~(\ref{dev1}) or  Eq.~(\ref{dev2}) by the order ${\cal O}(0.01)$. These explicitly show that the maximal value can increase, if the number of $m$ increases. However additional fields have to be introduced to make exotic fields decay into the SM fields appropriately in case where $4\le m$. 
The red scattering plots represents the $\mu-e$ conversion rate $R_{\rm Ti}$ in terms of the exotic fermion ($\Psi^\pm$) mass , where the upper figure represents $m=2$, and the lower one does $m=3$. Here the neutrino oscillation data and the LFVs as well as the current upper bound of $R_{\rm Ti}\lesssim 4.3\times 10^{-12}$ are taken into consideration as constraints.  Both of the minimal values of $R_{\rm Ti}$ is ${\cal O}(10^{-16})$, which will be tested in the future experiment that will reaches ${\cal O}(10^{-18})$. 
}
\label{m=2-3-damu}
\end{center}
\end{figure}

\subsection{The case of $m=2$ }
We show numerical results  for $m=2$ case as shown in the upper figures of ~\ref{m=2-3-damu}.
The upper-left side figure represents the scattering plot in the $M_{\Psi^\pm}$ mass and the muon anomalous magnetic moment,
and it suggests that the maximal discrepancy is $\Delta a_\mu={\cal O}(1.5 \times 10^{-11})$ that is below the experimental value ${\cal O}(10^{-9})$ in  Eq.~(\ref{dev1}) or  Eq.~(\ref{dev2}). However positive value can be obtained,
since the mass of  ${\Psi^\pm}$ cannot be a DM candidate. 
The upper-right side figure represents the scattering plot in the DM mass and the $\mu-e$ conversion rate with the target Ti that satisfies the current upper bound. Our minimal value of $R_{\rm Ti}$ is ${\cal O}(10^{-15})$, which can also be tested in the future experiment that will reaches ${\cal O}(10^{-18})$.
Now it is worthwhile about mentioning the way to make a decay for the exotic charged fields.
$\Psi^{\pm\pm} $ can always decay into the $W^\pm$ boson plus $\Psi^{\pm}$ because of $SU(2)_L$ doublet.
Then $\Psi^{2\pm} $ can decay into the singly charged boson $S^\pm$ plus neutrinos through the terms $y_L$ or $y_\Delta$. Finally $S^\pm$ can decay into the charged leptons and the neutrinos through the terms $y_L$ or $y_\Delta$ too.Totally the series of  process is as follows:
\begin{align}
\Psi^{2\pm}\to \Psi^{\pm}(+W^\pm) \to S^\pm(+\nu_L) \to \ell^\pm+\nu_L.
\end{align}
Thus all the exotic fields can decay into the SM fields appropriately even without the DM candidate in the case of $m=2$.

\subsection{The case of $m=3$}
We show numerical results  for $m=3$ case as shown in the lower  figures of ~\ref{m=2-3-damu}.
The lower-left side figure represents the scattering plot in the $M_{\Psi^\pm}$ mass and the muon anomalous magnetic moment,
and it suggests that the maximal discrepancy is $\Delta a_\mu={\cal O}(2\times 10^{-11})$ that is still below the experimental value ${\cal O}(10^{-9})$ in  Eq.~(\ref{dev1}) or  Eq.~(\ref{dev2}). However it increases comparing to the $m=2$ case by 1.5 times. 
The lower-right side figure represents the scattering plot in the DM mass and the $\mu-e$ conversion rate with the target Ti that satisfies the current upper bound. Our minimal value of $R_{\rm Ti}$ is ${\cal O}(10^{-15})$, which can also be tested in the future experiment that will reaches ${\cal O}(10^{-18})$.
The decay process is following.
$\Psi^{3\pm} $ can always decay into the $W^\pm$ boson plus $\Psi^{2\pm}$ because of $SU(2)_L$ doublet.
Then $\Psi^{\pm\pm} $ can decay into the doubly charged boson $S^{2\pm}$ plus neutrinos through the terms $y_L$ or $y_\Delta$. Finally $S^{2\pm}$ can decay into charged leptons with the same sign, because a new term $S^{++}\bar e_R^c e_R$ can be added into our Lagrangian. Totally the series of  process is as follows:
\begin{align}
\Psi^{3\pm}\to \Psi^{2\pm}(+W^\pm) \to S^{2\pm}(+\nu_L) \to 2\ell^\pm.
\end{align}
Thus all the exotic fields can decay into the SM fields appropriately  without introducing any kind of additional fields in the case of $m=3$.

\subsection{The case of $4\le m$}
In the case of $4\le m$, the value of muon $g-2$ can be obtained sizably, but we have to introduce new additional fields
in order to make the exotic fields decay into the DM fields.
For example, another singly boson ($h^\pm$) and doubly boson ($k^{2\pm}$) have to be introduced in the case of $m=4$. Then the additional terms are written as $k^{++}\bar e_R^c e_R$ and $\bar L_L^c L_L h^+$. Thus the following decay process can occur:
\begin{align}
S^{3\pm}\to k^{2\pm}+h^\pm \to 3\ell^{\pm}+\nu_L.
\end{align}

The case of $m=5$, another doubly charged boson $k^{2\pm}$ has to be introduced. Then the following decay process can occur:
\begin{align}
S^{4\pm}\to 2k^{2\pm} \to 4\ell^{\pm},\end{align}
due to the new term of $k^{++}\bar e_R^c e_R$.

\section{ Conclusions and discussions}
We have studied a one-loop induced radiative neutrino model with an inert isospin triplet boson in the general framework of $U(1)_Y$, in which we have discussed
neutrino oscillations, lepton flavor violations, anomalous magnetic moment, and a (singlet-like) bosonic dark matter candidate to explain the relic density and the direct detection (in case of $m=1$). Then we have found the several results from the global numerical analysis 
as follows. Firstly our maximal discrepancy for the muon anomalous magnetic moment are respectively given by $|\Delta a_\mu|={\cal O}(10^{-12})$ with negative value for $m=1$,  $\Delta a_\mu={\cal O}(1.5\times 10^{-11})$ for  $m=2$,  $\Delta a_\mu={\cal O}(2\times10^{-11})$ for  $m=3$, where the negative sign for $m=1$ directly comes from the constraint of the $\mu-e$ conversion.
Secondly the allowed region of DM mass ($m=1$) should be  around 
$1.7\ [{\rm TeV}] \le M_X\le 2.6\ [{\rm TeV}]$ except the resonant point, 
 which is rather narrow comparing to our scanning  space, whose result directly stems from the constraint of the relic density. Thirdly, the minimal values of $R_{\rm Ti}$ are ${\cal O}(10^{-16})$ for $m=1$ and ${\cal O}(10^{-15})$ for $m=(2-3)$, which will be tested in the future experiment that will reaches ${\cal O}(10^{-18})$. 
 When $4\le m$, we can obtain the sizable muon $g-2$, but we have found that some additional fields has to be introduced in order to make them decay into the SM fields.

The signal of our isospin triplet boson that might be measured at LHC potentially possesses the discrimination from the other radiative seesaw models, since few models introduce the triplet boson.
%
 The boson decays into the same sign dilepton + missing($=$DM) through the exotic leptons, and their corresponding Yukawa couplings $y_\Delta$ can always be larger than the typical values of Yukawa couplings in the type-II seesaw model because there are not any LFV processes at the tree level. It can occur when the DM mass  is the resonance point at around the half mass of the SM Higgs boson, which has been discussed in ref.~\cite{Kajiyama:2013zla}. 
%

\section*{Acknowledgments}
\vspace{0.5cm}
H. O. is sincerely grateful for all the KIAS members, Korean cordial persons, foods, culture, weather, and all the other things.
This work is supported in part by NRF Research No. 2009-0083526 (Y. O.) of the Republic of Korea.

\end{document}